\documentclass{epl}

\title{Casimir force between two ideal-conductor walls revisited}
\shorttitle{Casimir force revisited}

\author{
B. Jancovici\inst{1}\thanks{E-mail: \email{Bernard.Jancovici@th.u-psud.fr}} 
\and L. {\v S}amaj\inst{1,2}\thanks{E-mail: \email{fyzimaes@savba.sk}}
}
\shortauthor{B. Jancovici \etal}

\institute{
  \inst{1} Laboratoire de Physique Th\'eorique, Universit\'e de
Paris-Sud - B\^atiment 210, 91405 Orsay Cedex, France
(Unit\'e Mixte de Recherche no. 8627 - CNRS) \\
  \inst{2} Institute of Physics, Slovak Academy of Sciences -
D\'ubravsk\'a cesta 9, 845 11 Bratislava, Slovakia
}

\begin{document}

\maketitle

\begin{abstract}
The high-temperature aspects of the Casimir force between two neutral 
conducting walls are studied.
The mathematical model of ``inert'' ideal-conductor walls, considered 
in the original formulations of the Casimir effect, is based on 
the universal properties of the electromagnetic radiation in the vacuum
between the conductors, with zero boundary conditions for the tangential 
components of the electric field on the walls.
This formulation seems to be in agreement with experiments on metallic
conductors at room temperature.
At high temperatures or large distances, at least, fluctuations of 
the electric field are present in the bulk and at the surface of 
a particle system forming the walls, even in the high-density limit:
``living'' ideal conductors.
This makes the enforcement of the inert boundary conditions inadequate.
Within a hierarchy of length scales, the high-temperature Casimir 
force is shown to be entirely determined by the thermal fluctuations in 
the conducting walls, modelled microscopically by classical Coulomb fluids 
in the Debye-H\"uckel regime. The semi-classical regime, in the framework of
quantum electrodynamics is studied in the companion letter by P.R.Buenzli and 
Ph.A.Martin~\cite{Buenzli1}.
\end{abstract}

This letter is related to the one by Buenzli and Martin~\cite{Buenzli1}.
For the sake of completeness, we cannot avoid repeating a few things.

Casimir showed in his famous paper~\cite{Casimir} that fluctuations of 
the electromagnetic field in vacuum can be detected and quantitatively 
estimated via the measurement of a macroscopic attractive force between 
two parallel {\em neutral} metallic plates; 
for a nice introduction to the Casimir effect see~\cite{Duplantier} 
and for an exhaustive review see~\cite{Bordag1}. 

Let us recall briefly, within the formalism of Ref.~\cite{Duplantier}, 
some aspects of the usual theory for plates considered as made of ideal 
conductors, which are relevant in view of the present letter. 
We consider the 3D Cartesian space of points ${\bm r}=(x,y,z)$ where
a vacuum is localized in the subspace 
$\Lambda = \{{\bm r}\vert x\in(-d/2,d/2);(y,z)\in R^2 \}$ between two 
ideal-conductor walls (thick slabs) at a distance $d$ from each other.   
The time-dependent electric ${\bf E}({\bm r},t)$ and magnetic 
${\bf B}({\bm r},t)$ fields in $\Lambda$ are the solutions of the Maxwell 
equations in vacuum, subject to the boundary conditions that the tangential 
components of the electric field vanish at the ideal-conductor walls 
$\partial\Lambda=\{{\bm r}\vert x=\pm d/2;(y,z)\in R^2 \}$:
\begin{equation} \label{1}
E_y({\bm r},t) = E_z({\bm r},t) = 0
\qquad \mbox{for ${\bm r}\in \partial\Lambda$.}
\end{equation}
Note that this mathematical definition of the ideal-conductor wall 
is based on macroscopic electrostatics: the electric field is considered 
to be zero, without any fluctuation, inside the walls which have no
microscopic structure and act only as fixing the instantaneous 
boundary conditions of type (\ref{1}). 
We shall call such a mathematical model of an ideal conductor 
``inert ideal-conductor''.
For each separate mode labeled by the wave number ${\bm k}=(k_x,k_y,k_z)$
with $k_x = \pi n_x/d$ $(n_x=0,1,2,\ldots)$ and polarization indices 
$\lambda=1,2$ (only one polarization is possible when $k_x=0$), 
the quantized energy spectrum of the electromagnetic field between 
the walls corresponds to that of an oscillator with the frequency 
$\omega_{\bm k} = c \vert {\bm k}\vert$ ($c$ is the velocity of light).
At zero temperature $T=0$, no photons are present and so each mode contributes
by the zero-point energy $\hbar \omega_{\bm k}/2$ 
where $\hbar$ is Planck's constant. 
The $d$-dependent part of the system ground-state energy leads to
the following attractive Casimir force per unit surface 
of one of the walls  
\begin{equation} \label{2}
f^0(d) = - \frac{\pi^2 \hbar c}{240 d^4}\,.
\end{equation}

At nonzero temperature $T>0$, all numbers of photons
are possible and each mode contributes by the free energy
of the thermalized harmonic oscillator.
The Casimir force then reads
\begin{equation} \label{3}
f^T(d) = - \frac{2}{\pi\beta} \sum_{n=0}^{\infty}\,'
\int_0^{\infty} \upd k_{\perp}\, k_{\perp} q_n
\left( {\rm e}^{2 d q_n}-1 \right)^{-1}\,,
\end{equation} 
where $\beta=1/(k_B T)$ is the inverse temperature, the prime
in the sum over $n=0,1,2,\ldots$ means that the $n=0$ term should be
multiplied by $1/2$, $k_{\perp}$ is the magnitude of a wave-vector 
component in the $(y,z)$-plane and $q_n^2 = k_{\perp}^2+\xi_n^2/c^2$ 
with $\xi_n=2\pi n/(\hbar\beta)$ being the Matsubara frequencies.
By a simple change of variables, formula (\ref{3}) can be rewritten 
as follows
\begin{equation} \label{4}
f^T(d) = - \frac{1}{4\pi\beta d^3} \sum_{n=0}^{\infty}\,'
\int_{n t}^{\infty} \upd y\, y^2 \frac{1}{{\rm e}^y-1}\,,
\end{equation}
where 
\begin{equation} \label{5}
t = \frac{4\pi d}{\hbar c \beta}
\end{equation}
is the dimensionless parameter which measures the ratio of the separation
between the conductor walls to the thermal wavelength of a photon.
The small values of $t$ correspond to low temperatures or small 
distances where quantum effects dominate.
Using the Euler-MacLaurin sum formula, one obtains from eq.~(\ref{4})
the small-$t$ expansion of the form
\begin{equation} \label{6}
f^T(d) = - \frac{\pi^2 \hbar c}{240 d^4} - 
\frac{\pi^2}{45 (\hbar c)^3 \beta^4} + \frac{1}{\beta d^3}
O({\rm e}^{-4\pi^2/t}) , \qquad t\to 0\,.
\end{equation}
It is interesting that the leading correction to the $T=0$ result
(\ref{2}) is negligible in the experiments which have been performed
at room temperature, see for example refs. \cite{Lamoreaux,Bressi}.
The experiments are in good agreement with (\ref{6}).
The large values of $t$ correspond to high temperatures or large
distances where the classical limit of quantum mechanics provides
an adequate system description.
In the large-$t$ limit, the $n=0$ term dominates in the sum (\ref{4}),
which implies the classical $\hbar$-independent leading behavior
\begin{equation} \label{7}
f^T(d) = - \frac{\zeta(3)}{4\pi\beta d^3} + \frac{1}{\beta d^3}
O({\rm e}^{-t})\,, \qquad t\to\infty\,.
\end{equation}
For the present time, the high-$t$ region is not accessible to 
experiments on metals. 
However, the high-temperature regime might be of interest for electrolytes.

Lifshitz~\cite{Lifshitz} considered the more general case of dielectric walls 
with a frequency-dependent dielectric permittivity $\epsilon(\omega)$.
His starting point was the fluctuations within the walls, which
therefore were not considered as inert.
He derived the following formula for the Casimir force~\cite{Bordag1}:
\begin{eqnarray} 
f^T(d) & = & - \frac{1}{\pi\beta} \sum_{n=0}^{\infty}\,'
\int_0^{\infty} \upd k_{\perp}\, k_{\perp} q_n \nonumber \\
& & \times \left\{ 
\left[ r_{\parallel}^{-2}(\xi_n,k_{\perp}) {\rm e}^{2 d q_n} -1 \right]^{-1}
+ \left[ r_{\perp}^{-2}(\xi_n,k_{\perp}) {\rm e}^{2 d q_n} -1 \right]^{-1}
\right\} , \label{8}
\end{eqnarray}
where $r_{\parallel}$ and $r_{\perp}$ are the reflection coefficients 
of the TM and TE modes, respectively.
They are given by
\begin{equation} \label{9}
r_{\parallel}^{-2}(\xi_n,k_{\perp}) = \left[ \frac{\epsilon({\rm i}\xi_n) q_n 
+ k_n}{\epsilon({\rm i}\xi_n) q_n - k_n} \right]^2\,, \qquad
r_{\perp}^{-2}(\xi_n,k_{\perp}) = \left( \frac{q_n + k_n}{q_n - k_n} 
\right)^2\,,
\end{equation}
with $k_n^2 = k_{\perp}^2 + \epsilon({\rm i}\xi_n) \xi_n^2/c^2$.
When $\epsilon(\omega)<\infty$, eqs.~(\ref{8}) and (\ref{9}) are well
defined.
When $\epsilon(\omega)\to\infty$, the zero-frequency $n=0$ term
in the sum on the rhs of (\ref{8}) is not uniquely defined because
its value depends on the order of the limits  
$\epsilon(\omega)\to\infty$ and $n\to 0$.
In order to restore the inert ideal-conductor result (\ref{3}) based
on the electrostatic boundary conditions (\ref{1}), 
Schwinger et al.~\cite{Schwinger} postulated the following order: 
set first $\epsilon(\omega)=\infty$, then take the limit $n=0$.
This prescription implies the reflection coefficients of the zero mode 
to be $r_{\parallel}^2(0,k_{\perp}) = r_{\perp}^2(0,k_{\perp}) =1$
for inert ideal metals.

Experiments are performed on real conductors composed of quantum particles,
with finite static conductivity $\sigma$ and plasma frequency $\omega_p$, 
given by $\omega_p^2 = 4\pi e^2 n/m$ where $n$ is the number density of 
free electrons of mass $m$.
For such real conductors, one has the Drude formulae for 
the frequency-dependent $\epsilon(\omega)$:
\begin{eqnarray}
\epsilon(\omega) & \sim & \frac{4\pi {\rm i} \sigma}{\omega}
\qquad \mbox{for $\omega\to 0$}\,, \label{10} \\
\epsilon(\omega) & \sim & 1 - \frac{\omega_p^2}{\omega^2}
\qquad \mbox{for $\omega \gg \omega_p^2/(4\pi\sigma)$}\,. \label{11} 
\end{eqnarray}  
The consideration of a frequency-dependent $\epsilon(\omega)$ enables
one to avoid an artificial prescription for the order of limits:
it is the dynamics of the particle system which ``chooses'' the
correct treatment of the zero-mode contribution.
In a series of recent works~\cite{Bostrom,Hoye1,Milton,Hoye2,Hoye3},
the Drude formula (\ref{10}) was substituted into eq.~(\ref{9})
considered for the zero Matsubara frequency $\xi_0\to 0$. 
This leads to the reflection coefficients
$r_{\parallel}^2(0,k_{\perp}) = 1, r_{\perp}^2(0,k_{\perp}) = 0$
independent of $\sigma$, i.e. for $n=0$ the second term on the rhs of 
eq.~(\ref{8}) does not contribute to the Casimir force for a real conductor.
As a mathematical consequence, the additional term $\zeta(3)/(8\pi\beta d^3)$
appears in the Casimir force in any regime.
In particular, the large-temperature formula (\ref{7}) is modified to
\begin{equation} \label{12}
f^T_{\mathrm{L}}(d)\sim - \frac{\zeta(3)}{8\pi\beta d^3} \qquad 
\mbox{for $t\to\infty$}\,,
\end{equation}
a result identical to the one given by Lifshitz~\cite{Lifshitz}.
Although the additional term vanishes at zero temperature, it is
relevant in the region of small temperatures where it is the source
of some contradictions.
Namely, it was argued in another series of 
works~\cite{Bordag2,Geyer,Klimchitskaya} that, at low temperatures, 
the relation (\ref{11}) should be used.
An intensive polemic about the low-temperature Casimir effect 
persists in our days~\cite{Hoye3,Klimchitskaya}.

In this letter, we shall concentrate on the high-temperature aspects of
the Casimir effect.
There is an apparent discrepancy by a factor $1/2$ between 
the high-temperature Schwinger formula (\ref{7}), valid for inert 
ideal-conductor walls with the boundary conditions (\ref{1}), 
and the Lifshitz formula (\ref{12}), valid for real-conductor walls 
with $\epsilon(\omega)$ given by the Drude dispersion relation (\ref{10}).
We aim at explaining this discrepancy on the basis of some exact results 
for specific microscopic particle systems which are used to model 
the conductor walls.  
The consideration of the Casimir effect in the $t\to\infty$ limit
is also motivated by two fundamental simplifications of these model systems.
First, according to the correspondence principle, in a microscopic model
of matter coupled to electromagnetic radiation at equilibrium, both matter 
and radiation can be treated classically in the high-temperature limit.
This fact manifests itself as the absence of $\hbar$ in the
leading terms of the expansions (\ref{7}) and (\ref{12}).
Second, the application of the Bohr-van Leeuwen theorem~\cite{Bohr,vanLeeuwen}
leads to the decoupling between classical matter and radiation, and to an
effective elimination of the magnetic forces in the matter
(for a nice detailed treatment of this subject, see ref.~\cite{Alastuey}).
The absence of relativistic effects is seen via the independence of 
the leading terms in eqs.~(\ref{7}) and (\ref{12}) on $c$.
We conclude that the matter can be treated in the $t\to\infty$ limit
as a classical matter, unaffected by radiation, where the charges interact
only via the instantaneous Coulomb potential.

As a model system of the classical Coulomb fluid, we consider a general
mixture of $M$ species of mobile pointlike (structureless) particles 
$\alpha = 1,2,\ldots$ with the corresponding masses $m_{\alpha}$ and 
charges $Z_{\alpha} e$, where $e$ is the elementary charge and $Z$ denotes 
integer valence ($Z=-1$ for an electron).
Its statistical mechanics is treated in the grand canonical
ensemble characterized by the inverse temperature $\beta$ and by the
species fugacities $\{ z_{\alpha}\}$ or, equivalently, the bulk species 
densities $\{ n_{\alpha} \}$ constrained by the neutrality condition
$\sum_{\alpha} Z_{\alpha} n_{\alpha} = 0$. 
The thermal average will be denoted by $\langle \cdots \rangle$.
We use Gaussian units.
The interaction energy of particles $\{ i \}$ with charges $\{ q_i \}$,
localized at spatial positions $\{ {\bm r}_i \}$, is given by
$\sum_{i<j} [ q_i q_j v(\vert {\bm r}_i-{\bm r}_j \vert) 
+ u_{(\lambda_i+\lambda_j)/2}(\vert {\bm r}_i-{\bm r}_j \vert)]$, 
where $v(r)=1/r$ is the Coulomb potential and
\begin{equation} \label{13} 
u_{\lambda}(r) = \left\{ 
\begin{array}{ll} 
\infty & \mbox{for $r<\lambda$}\,, \\
0 & \mbox{for $r\ge \lambda$}\,, 
\end{array} \right.
\end{equation}
is the hard-core repulsion potential which prevents the classical 
thermodynamic collapse between oppositely charged particles.
To make the correspondence with the quantum-mechanical version of
the model, the hard-core diameter of particles of type $\alpha$ 
has to be set equal to the thermal de Broglie wavelength 
$\lambda_{\alpha} = \hbar (2\pi\beta/m_{\alpha})^{1/2}$~\cite{Brydges}.

We would like to emphasize that the present particle system represents
a microscopic model of ``living conductors'' where the charge density 
and the corresponding electric potential/field fluctuate, even 
for extreme values of physical parameters like the particle density.
To be more precise, let us consider the truncated charge-charge 
correlation function
\begin{equation} \label{14}
S({\bm r},{\bm r}') = 
\langle {\hat\rho}({\bm r}) {\hat\rho}({\bm r}') \rangle^{\mathrm{T}}\,,
\end{equation}  
where the microscopic charge density ${\hat\rho}$ is defined by
${\hat\rho}({\bm r}) = \sum_{\alpha} Z_{\alpha} e {\hat n}_{\alpha}({\bm r})$
with ${\hat n}_{\alpha}({\bm r}) = \sum_i \delta(Z_{\alpha}e,q_i) 
\delta({\bm r}-{\bm r}_i)$ being the microscopic number density of 
$\alpha$-species.
In the infinite space (bulk regime), the fact that the Fourier transform 
of the Coulomb interaction has the form ${\tilde v}({\bm k}) = 4\pi/k^2$
implies the following small-$k$ behavior of the charge structure function
(the Fourier transform of (\ref{14}) with respect to
$\vert {\bm r}-{\bm r}'\vert$)
\begin{equation} \label{15}
{\tilde S}(k) = \frac{1}{4\pi\beta} k^2 + O(k^4)\,;
\end{equation}
for a review of sum rules for charged systems, see ref.~\cite{Martin}. 
This exact result is not influenced by short-range interaction potentials
like the hard-core one.
Thus, the second moment of $S(r)$ {\em does not} depend on the total
particle number density 
$n=\sum_{\alpha} \langle {\hat n}_{\alpha}({\bm r}) \rangle$
and survives also in the high-density region.
An immediate consequence of eq.~(\ref{15}) is an asymptotic formula
for the long-ranged potential-potential correlation function
\cite{Lebowitz}:
\begin{equation} \label{16}
\beta \langle {\hat\phi}({\bm r}) {\hat\phi}({\bm r}') 
\rangle^{\mathrm{T}} \sim \frac{1}{\vert {\bm r}-{\bm r}'\vert}
\qquad \mbox{as $\vert {\bm r}-{\bm r}' \vert \to \infty$}\,.
\end{equation}
Here, ${\hat\phi}({\bm r}) = \int \upd {\bm r}'\, 
v(\vert {\bm r}-{\bm r}'\vert) {\hat\rho}({\bm r}')$ is the
microscopic electric potential created at point ${\bm r}$ by
the system of charged particles, and the distance 
$\vert {\bm r}-{\bm r}'\vert$ has to be large compared to the
microscopic scale represented by the correlation length of the
short-ranged (exponentially decaying) particle correlations.
Since the microscopic electric field ${\hat{\bf E}}$ is given by
${\hat E}_{\mu}({\bm r}) = - \partial_{\mu} {\hat\phi}({\bm r})$
$(\mu=x,y,z)$, the field-field correlation function is obtained
from eq.~(\ref{16}) as
\begin{equation} \label{17}
\beta \langle {\hat E}_{\mu}({\bm r}) {\hat E}_{\nu}({\bm r}')
\rangle^{\mathrm{T}} \sim
\frac{3({\bm r}-{\bm r}')_{\mu} ({\bm r}-{\bm r}')_{\nu}
- \delta_{\mu\nu} \vert {\bm r}-{\bm r}'\vert^2}{
\vert {\bm r}-{\bm r}' \vert^5}\,.
\end{equation}  
It is obvious that $\langle {\hat{\bf E}}({\bm r}) \rangle = 0$
since the mean electric potential is a constant inside a conductor.
However, the asymptotic formula (\ref{17}) tells us that nonzero
thermal fluctuations of the electric field must be present in 
the system for any particle density $n$.
The generalization of the fluctuation results, obtained for the bulk, 
to inhomogeneous situations of the present interest, like conductors
with boundaries, was made in ref.~\cite{Jancovici95}.
As soon as the two points ${\bm r}$ and ${\bm r}'$ are inside
a conductor, asymptotic formulae (\ref{16}) and (\ref{17}) remain valid.
When one of the points lies on the conductor boundary, the tangential
components of the electric field at this point still fluctuate 
according to (\ref{17}), while the discontinuity of the normal component
across the surface is related to surface charge fluctuations.
These fluctuation phenomena make the living conductors fundamentally
different from the inert ones with tangential components of 
the electric field at a boundary identically set to zero,
as in eq.~(\ref{1}). 
The Casimir force (\ref{12}) can be retrieved through a Maxwell stress 
tensor computed from the electric-field fluctuations 
in the vacuum region~\cite{Jancovici04}.

As was already mentioned, our Coulomb fluid of classical charged particles 
with de Broglie hard cores can represent its quantum counterpart of
pointlike charges at sufficiently large temperatures. It has been shown in
~\cite{Buenzli1} that the long-range charge correlations of the semi-classical
regime do not spoil the classical limit (\ref{12}).
The high-temperature region of classical fluids is described exactly
by the Debye-H\"uckel (DH) theory.
Rigorous conditions, under which the DH approximation gives the exact 
leading correction to the ideal gas, were the subject of many studies 
in the past; for a short historical review, see e.g.~\cite{Brydges}.
These conditions arise naturally in a renormalized Mayer diagrammatic
expansion for statistical quantities~\cite{Aqua,Cornu}.
In terms of the mean interparticle distance $a$ and the inverse Debye 
length $\kappa$ ($\kappa^{-1}$ is the correlation length of particles
in the DH regime), defined by
\begin{equation} \label{18}
\frac{4\pi a^3}{3} = \frac{1}{n}\,, \qquad
\kappa^2 = 4\pi \beta e^2 \sum_{\alpha} Z_{\alpha}^2 n_{\alpha}\,,
\end{equation}
the DH scaling regime is given by (see Eq. (11) of ref.~\cite{Cornu})
\begin{equation} \label{19}
\left( \frac{\lambda}{a} \right)^3 \ll
\frac{1}{2} \kappa \beta e^2 \ll 1\,,
\end{equation}
where $\lambda$ represents the ``typical'', in our case de Broglie,
hard-core radius of particles.
Supposing in what follows for the sake of simplicity that 
$\sum_{\alpha} Z_{\alpha}^2 n_{\alpha}/n$ is of order of unity and 
omitting irrelevant numerical factors, these inequalities can be 
rewritten in a more transparent form
\begin{equation} \label{20}
a_0 \ll a \ll \kappa^{-1}
\end{equation}  
where $a_0 \sim \hbar^2/(m e^2)$ with $m$ being the ``typical'' particle mass. 
The lightest of the charges are the electrons for which the quantum
microscopic scale $a_0$ attains its maximum value, equal to 
the Bohr radius $\sim 10^{-10}$m. 
The first ``Bohr'' inequality in (\ref{20}) is a quantum
upper bound for possible values of particle densities. 
Since the Bohr radius is small, very dense Coulomb fluids
with $a\sim 10^{-8}$m are allowed; we shall refer to them as 
``living ideal conductors''.
The second inequality combines both the particle density $n$ and 
the temperature parameter $\beta e^2$; for a fixed particle density
allowed by the first inequality, there always exists a sufficiently 
high temperature above which this inequality is fulfilled.

Let our classical Coulomb fluid model the conductor slabs in the
Casimir geometry.
The characteristic correlation length of the particle system 
is assumed to be much smaller than the macroscopic vacuum distance 
$d$ between the conductor walls,
\begin{equation} \label{21}
\kappa^{-1} \ll d\,.
\end{equation}
Then, the $t$-parameter (\ref{5}) is of the form $A (\kappa d)/(\kappa a)^3$ 
with $A$ of the order of $12 \pi e^2/(\hbar c)$, i.e. unity. 
The scale hierarchy (\ref{20}) and (\ref{21}) is thus fully consistent 
with the classical and nonrelativistic limit of interest $t\to\infty$.
To summarize: as soon as the scaling length regimes (\ref{20}) and
(\ref{21}) apply, the quantum system of charged particles coupled 
to electromagnetic radiation at equilibrium can be represented 
in terms of its classical pure-Coulomb fluid counterpart, decoupled
from radiation and treated within the DH theory.
The Casimir force originates exclusively from the thermal
fluctuations in the conducting walls modelled by
this classical Coulomb fluid.

The Casimir problem of microscopic Coulomb fluids was solved by using 
an inhomogeneous version of the DH theory in two recent papers: 
the work \cite{Jancovici04} dealt also with more complex physical situations,
the study \cite{Buenzli2} went beyond the DH theory.
In the DH theory, the large-$\kappa d$ expansion of the Casimir force 
was obtained in the form \cite{Jancovici04}
\begin{equation} \label{22}
f^{\mathrm{T}}(d) = - \frac{\zeta(3)}{8\pi\beta d^3}
\left\{ 1 - \frac{6}{(\kappa d)} + 
O\left( \frac{1}{(\kappa d)^2} \right) \right\}\,.
\end{equation}
The leading universal term is identical to the Lifshitz result (\ref{12}).
The subleading correction term is non-universal and depends on 
the composition of the Coulomb fluid via $\kappa$.  
Even for a very dense Coulomb fluid with the mean interparticle distance
$a\sim 10^{-8}$m (living ideal conductor), there exists a sufficiently  
high temperature and a sufficiently large distance between slabs 
above which the required length scale hierarchy $a\ll \kappa^{-1}\ll d$ 
takes place and the correction term is negligibly small in comparison 
with the leading one. 

In conclusion, the mathematical model of inert ideal-conductor
walls is based on the zero boundary conditions for 
the tangential components of the electric field (\ref{1}).
This seems to be in agreement with experimental results at zero temperature, 
and perhaps also at sufficiently small temperatures.
Why the quantum ground-state fluctuations in the real walls
seem to play no role is an open problem.
At high temperatures, fluctuations of the electric field prevail in 
the bulk and at the surface of the particle system, even in 
the high-density limit (living ideal conductor), which makes 
the enforcement of the inert boundary conditions inadequate.
Within the hierarchy of length scales (\ref{20}) and (\ref{21}), 
the high-temperature Casimir force was shown to be entirely 
determined by the thermal fluctuations of the conducting walls, 
modelled microscopically by classical Coulomb fluids 
in the Debye-H\"uckel regime.

\acknowledgments
We thank Ph. A. Martin and P. Buenzli for very fruitful discussions
on the subject of this work.
The authors acknowledge support from the CNRS-SAS agreement.
A partial support of L. \v Samaj by a VEGA grant is acknowledged.

\end{document}